\documentclass[aps,pre,twocolumn,reprint,preprintnumbers,floatfix,nofootinbib,superscriptaddress]{revtex4-1}

\usepackage{amsmath}
\usepackage{amssymb}

\usepackage{graphicx}
\usepackage{subcaption}
\usepackage{dcolumn}
\usepackage{bm}
\usepackage{marvosym}
\usepackage{epstopdf} 
 
\usepackage{color}
\definecolor{light-gray}{gray}{0.5}
\definecolor{blue}{rgb}{0.0,0.0,1.0}
\definecolor{green}{rgb}{0.0,0.5,0.0}
\definecolor{red}{rgb}{1.0,0.0,0.0}
\definecolor{cyan}{rgb}{0.0,0.75,0.75}
\definecolor{magenta}{rgb}{0.75,0.0,0.75}
\definecolor{yellow}{rgb}{0.75,0.75,0.0}

\newcommand{\avg}[1]{\langle{#1}\rangle}
\newcommand{\sdot}{\cdot}

\newcommand{\grad}{\bm \nabla}
\newcommand{\pd}{\partial}

\begin{document}
% \title{Energy flux and dissipation rate in decaying rotating turbulence}
% \title{\vd{Spectral imbalance in decaying rotating turbulence}}
% \title{\vd{Spectral imbalance in the interscale dynamics of decaying rotating turbulence}}
% \title{\vd{Spectral imbalance between energy flux and dissipation in decaying rotating turbulence}}
\title{Spectral imbalance in the inertial range dynamics of decaying rotating turbulence}

\author{P. Valente}
% \email{}
\affiliation{LAETA/IDMEC, Instituto Superior Tecnico, Universidade de Lisboa, Av. Rovisco Pais, 1049-001 Lisboa, Portugal}
\author{V. Dallas}
% \email{}
\affiliation{Department of Applied Mathematics, University of Leeds, Leeds LS2 9JT, UK}

\begin{abstract}
Direct numerical simulations of homogeneous decaying turbulence with mild background rotation show the existence of a systematic and significant imbalance between the non-linear energy cascade to small scales and its dissipation. By starting the decay from a statistically stationary and fully developed rotating turbulence state, where the dissipation and the energy flux are approximately equal, the data shows a growing imbalance between the two until a maximum is reached when the dissipation is about twice the energy flux. This dichotomy of behaviours during decay is reminiscent of the non-equilibrium and the equilibrium regions previously reported for non-rotating turbulence [P.C. Valente, J.C. Vassilicos, Phys. Rev. Lett. {\bf 108} 214503 (2012)]. Note, however, that for decaying rotating turbulence the classical scaling of the dissipation rate $\epsilon \propto u'^3/L$ (where $u'$ and $L$ are the root mean square fluctuating velocity and the integral length scale, respectively) does not appear to hold during decay, which may be attributed to the effect of the background rotation on the energy cascade. On the other hand, the maximum energy flux holds the scaling $\Pi_{max} \propto u'^3/L$ in the initial stage of the decay until the maximum imbalance is reached.
\end{abstract}

\maketitle

\section{\label{sec:intro}Introduction}
Estimating the small-scale energy dissipation from large scale statistics for statistical stationary and non-stationary evolving flows is at the core of virtually all turbulence closures as it enables predictions of momentum transport (e.g., drag), mixing, particle dispersion/clustering, noise, etc.
The emphasis is given to the empirical scaling of the energy dissipation rate, $\varepsilon \propto u'^3/L$,
using solely a characteristic turbulent velocity, such as the root mean square of the velocity fluctuations $u'$
% Pedro: can we not be more specific, i.e. the rms velocity?
and an integral length scale $L$, is directly related to its practical application for modelling and goes back to the %original
seminal works by G.I. Taylor \cite{Taylor1935,*Taylor1935b,*Taylor1935c} and A.N. Kolmogorov \cite{K41a,*K41b,*K41c,*K42}.
This inviscid scaling of the viscous dissipation rate of energy is supported by the widely accepted phenomenology that the small-scale dissipative turbulence-induced motions are fed by a continuous range of larger-scale motions (the energy cascade \cite{Frisch:book,MY2010}), and are always sufficiently small-scale to make molecular %transport/
dissipation efficient (also known as the dissipation anomaly \cite{Eyink2003,Frisch:book}). 

%To distinguish between exact and empirical forms of $4/5^{\mathrm{th}}$-type laws, it is instructive to summarise the derivation of the exact law which can be done either in physical or wavenumber space \cite{NT99,Cambon2012}.
The scale-by-scale energy budget for incompressible, externally forced, homogeneous flows in wavenumber space can be written as \cite{Frisch:book,lin1947remarks},
\begin{equation}\label{lin_1st} 
 \partial_t E(k,t) = T(k,t) - 2\nu k^2 E(k,t) + F(k,t),
\end{equation} 
where $E(k,t)$ and $T(k,t)$ are, respectively, the spherically averaged energy spectrum and the net energy %exchanged 
transfer term, %from wavenumber $k=|{\bf k}|$ with other wavenumbers, 
$2\nu k^2 E(k)$ is the viscous dissipation spectrum and $F(k,t)$ is the spectrum of energy input from the external forcing. 
%Note that anisotropy is dealt with by removing the directional dependence with the spherical averaging.  
Note that the energy budget is the same both for rotating and non-rotating flows \cite{lin1947remarks}.
Supposing that the external forcing is concentrated %at the large scales %corresponding to wavenumbers smaller than $k_f$ 
at small wavenumber $k_f$ and integrating each term in Eq. \eqref{lin_1st} from $k>k_f$ to $\infty$ we get,
\begin{equation}
\int_k^{\infty} \!\!\partial_t E(k',t)\, dk' = \Pi(k) - \int_k^{\infty} \!\!2\nu k'^2 E(k',t)\, dk', 
  \label{eq3}
 \end{equation}
 where $\Pi(k,t) \equiv \int_k^{\infty} T(k',t) dk'$ is the non-linear energy flux.  
% Now, it is possible to motivate, and 
It is generally accepted, that for large Reynolds numbers and for $k$ within the inertial range of scales $\int_k^{\infty} 2\nu k'^2 E(k',t) dk' \approx \varepsilon$ 
(i.e. the contribution of the large scales to the viscous dissipation is negligible) \cite{Cambon2012,VV2015}. 
 Here, the inertial range corresponds to scales sufficiently small not to have external energy input, i.e. $k>k_f$, but large enough for their contribution to the viscous dissipation to be negligible. 
Note that one cannot neglect $\int_k^{\infty} \partial_t E(k',t) \, dk'$ without introducing Kolmogorov's notion of local equilibrium \cite{K41c} or restricting the scope to statistically steady turbulence where this term is identically zero and thus $\Pi(k) \approx \varepsilon$ for any $k$ within the inertial range of scales, as long as the contribution to the viscous dissipation from $\int_0^{k} 2\nu k'^2 E(k',t) dk'$ is negligible, which is considered to be asymptotically exact for infinite Reynolds numbers.
Kolmogorov's notion of local equilibrium assumes that small-scale turbulent motions are very fast-paced and thus instantaneously adjust to dissipate whatever energy they are fed. The conceived near-instantaneous adjustment of the level of dissipation to the energy that the small scales receive from the large scales via the non-linear flux (i.e. $\Pi(k) \approx \varepsilon$), is a landmark of the classical theory of turbulence and became popularised as Kolmogorov's $4/5^{\mathrm{th}}$ law due to its isotropic form \cite{Frisch:book, Pope, MeneveauKatz}. 
The generalisation that $\int_k^{\infty} \partial_t E(k',t) \, dk'  \approx 0$ and therefore $\Pi(k) \approx \varepsilon$ for virtually all turbulent flows justifies its importance for turbulence modelling and its use as a building block in state-of-the-art closures.
%%
% Furthermore, even for statistically steady flows, the balance $\Pi(k) = \varepsilon$ is exact in a statistical sense where the quantities $\Pi(k) $ and $ \varepsilon$ are taken as averages in time and in the homogeneous directions and it is known not to hold in a local sense, at least for finite Reynolds \cite{Kraichnan74,BO98,Pearson04}.}
 
Consequently, for %temporally evolving flows (in either a Eulerian or Lagrangian frame of reference)
decaying or generally non-stationary flows the spectral balance $\Pi(k) \approx \varepsilon$ (or $\Pi(k) = \varepsilon$ at infinitely large Reynolds numbers) is not exact and requires empirical testing to support its use in turbulence closures.
Even for statistically steady flows at very large Reynolds numbers, the balance $\Pi(k) = \varepsilon$ is exact in a statistical sense where the quantities $\Pi(k) $ and $ \varepsilon$ are taken as averages in time and in the homogeneous directions and it is known not to hold in a local sense
%, at least for finite Reynolds 
\cite{Kraichnan74,BO98,Pearson04}.
However, for non-stationary flows the balance $\Pi(k) = \varepsilon$ is yet to be observed - a fact that is usually attributed either to  the data being at insufficiently high Reynolds numbers (i.e. a low Reynolds number effect \cite{Antonia2006,Cambon2012}) or, contrastingly, a consequence of the delay in cascading the energy down to the small-scales (a lag which increases with the Reynolds number \cite{Kraichnan74,Lumley92B,Yoshizawa1994,Rubinstein2003,Bos2007}).

A third, alternative viewpoint is that, regardless of the Reynolds number and/or of energy cascade `delays', one cannot neglect the required rate of change of energy to induce or annihilate small scale motions $\int_k^{\infty} \pd_t E(k',t) \, dk'$ 
%\red{to induce or annihilate fine scale motions} 
\cite{VOS2014}. 
% Pedro: I suppose you state this based on the balance from Eq. 1.
%
This is argued to be the case because even though the fraction of the total energy contained in the small-scales $K_{\eta}$ decreases with the Reynolds numbers, the associated time-scale $\tau_{\eta}$ also becomes vanishingly small and it can be  shown that %their ratio is proportional to the dissipation itself 
$K_{\eta}/\tau_{\eta} \propto \varepsilon$ and thus finite.
This can be argued to be the root cause for the significant imbalance between $\Pi(k)$ and $\varepsilon$ reported for non-stationary homogeneous turbulence and the manifestations of non-equilibrium dissipation behaviour observed in recent experiments and simulations \cite{nagataetal08a,GGL2010,VV2012,Nagata2012,VV2015,VOS2014}.

Many of the considerations above are also applicable for mildly rotating turbulent flows, i.e. flows where rotation has an important role on turbulence dynamics but it does not fully dominate the flow and lead to a quasi-2D %weak 
turbulence regime \cite{Bartelo2007,mininni09,Cambon2014,dt16,Alexakis2015}.
For example, the empirical scaling %for the energy dissipation rate 
$\varepsilon \propto u'^3/L$ is thought to apply to mildly rotating turbulence \cite{Alexakis2015} with different variants to take into account the anisotropy of the flow \cite{Squires1994,Davidson2015}.
% 
%Also, Eqs. \ref{lin_1st} and \ref{eq3} are applicable to turbulence with background rotation  \cite{Galtier2009,Moisy2011}, although the derived balance $\Pi(k) \approx \varepsilon$ also requires Kolmogorov's local equilibrium hypothesis to be applicable to temporally evolving flows.
%
In contrast, strongly rotating flows are weakly turbulent and exhibit marked differences such as laminar-like dissipation scaling \cite{Alexakis2015} and thus the discussions pertaining to turbulence theory are of limited use.
Furthermore, this mild rotation regime has, arguably, a closer connection to engineering applications and is also typical of many rotating turbulence laboratory experiments and numerical simulations
\cite{Cambon1990,Moisy2005,ruppertetal05,Bartelo2007,davidsonetal12ten,Alexakis2015,godeferdmoisy15}.
% Shall we mention that large-scale motions in the atmosphere, such as the Jet Stream and synoptic storms typically have moderate Rossby numbers, i.e. $Ro \leq 0.5$ \cite{mcwilliamsGFDbook06}

% \red{Given that mildly rotating turbulent flows are not excessively dissimilar from non-rotating turbulent flows,} it may be expected that the recent findings of non-equilibrium dissipation scalings and significant imbalances between \vd{$\Pi(k)$} and $\varepsilon$ for decaying non-rotating turbulence may also occur in mildly rotating decaying flows.
In this paper, we investigate the existence of significant imbalances between $\Pi(k)$ and $\varepsilon$ and non-equilibrium dissipation scalings in mildly rotating decaying flows.
%
%To explore this possibility, 
Therefore, we perform Direct Numerical Simulations (DNS) of decaying periodic box turbulence subject to different background rotation rates $\Omega$ and we create the conditions for non-equilibrium dissipation scalings by using a statistically steady and fully-developed rotating turbulence field as an initial condition \cite{VOS2014}.

\section{\label{sec:dns}Numerical methodology}
In this study, we consider the three-dimensional (3D) incompressible Navier-Stokes equations in a rotating frame of reference
\begin{equation}
 \pd_t \bm u + \bm \omega \times \bm u + 2\bm \Omega \times \bm u = - \grad P + \nu \grad^2 \bm u,
 \label{eq:ns}
\end{equation}
where $\bm u$ is the velocity field, $\bm \omega = \grad \times \bm u$ is the vorticity, $P$ is the pressure and $\nu$ is the kinematic viscosity. In a Cartesian domain, we choose the rotation axis to be in the $z$ direction with $\bm \Omega = \Omega \bm e_z$, where $\Omega$ is the rotation frequency. In the ideal case of $\nu = 0$, Eq. \eqref{eq:ns} conserves the energy $E = \frac{1}{2}\avg{|\bm u|^2}$ (where $|\sdot|$ stands for the $L_2$-norm) and the helicity $H = \avg{\bm u \sdot \bm \omega}$ with the angular brackets denoting a spatial average. %unless indicated otherwise. 

We numerically integrate Eq. \eqref{eq:ns} using the pseudo-spectral method in a periodic box of size $2\pi$ satisfying the incompressibility condition $\grad \sdot \bm u = 0$ and using a third-order Runge-Kutta scheme for the temporal advancement. The aliasing errors are removed with the $2/3$ dealiasing rule and as a result the minimum and maximum wavenumbers are $k_{min}=1$ and $k_{max}=N/3$, respectively, where $N$ is the number of grid points in each Cartesian coordinate. For more details on the numerical code, see Ref. \citep{mpicode05}.

The initial conditions for the decaying simulations are obtained by running the code with an additional non-helical random forcing term (see \cite{dfa15,dt16}) until a statistically steady and fully developed turbulence state is reached. All simulations were integrated for more than 100 turnover times with the exception of the highest resolution runs ($1024^3$), which were integrated for roughly 80 turnover times. Then, the free turbulence decay was initiated by switching off the forcing.

%\textcolor{magenta}{The simulations are benchmarked against data obtained with an independent numerical code using the same numerical scheme (see Ref. \cite{VOS2014}) but a different forcing strategy. The main difference between the two non-helical random forcing strategies is that the latter is an impulsive force uncorrelated with the velocity field and thus allows to set \emph{a priori} the level of dissipation in the steady state, whereas the former specifies the amplitude of the forcing.}

%We extract 
The turbulent %kinetic 
energy $K$, the energy dissipation rate $\varepsilon$ and the integral scale $L$ are extracted from the spherical-shell averaged %kinetic 
energy spectrum $E(k) \equiv \sum_{k \leq |\bm k| < k+1} |\widehat{\bm u}_{\bm k}|^2$, as 
% $K \equiv \sum_{k} E(k)$,  
% $\varepsilon \equiv 2 \nu \sum_k k^2 E(k)$ and 
% $L \equiv 3\pi/(4 K) \sum_{k} E(k)/k$, 
\begin{align}
 K &\equiv \sum_{k} E(k) \\
 \varepsilon &\equiv 2 \nu \sum_k k^2 E(k) \\
 L &\equiv 3\pi/(4 K) \sum_{k} E(k)/k
\end{align}
where $\widehat{.}$ denotes the Fourier mode.
%
%The spectrum is computed as, $E(k)=2 \pi k^2 \left< \frac{1}{2 } \widehat{u}_i (\vec{k},t) \widehat{u}_i^{\ast} (\vec{k},t) \right>_{|\vec{k}|}$, where $\widehat{u}_i(\vec{k},t)=FFT \Big\{ u_i (\vec{x},t) \Big\}$ is the Fourier coefficient of the velocity field $u_i (\vec{x},t)$ and `$^\ast$' represents a complex conjugate ($FFT$ is the direct Fourier transform operation), while $\left< \widehat{\phi} (\vec{k},t) \right>_{|\vec{k}|}=\frac{1}{N_k} \sum_{k-\frac{\Delta k}{2}<|\vec{k}|<k+\frac{\Delta k}{2}} \widehat{\phi} (\vec{k},t)$, denotes a spherical-shell average of the quantity $\widehat{\phi}$ over $N_k$ modes in the shell, of thickness $\Delta k$, centred at $k$.
%
%The energy cascade at wavenumber $k$ is computed from the non-linear energy transfer $T(k)$ as $\Pi(k) \equiv - \sum_{k''<=k} T(k'')$, from which we compute the maximum downscale energy flux $\Pi_{\max} \equiv \max[\Pi(k)]$ (note that the method for obtaining $E(k)$ and $T(k)$ from the velocity field can be found in Refs. \cite{Frisch:book, Pope}). 
The energy flux at wavenumber $k$ is computed 
% from the non-linear energy transfer 
% $T(k) \equiv \sum_{k \leq |\bm k| < k+1} \widehat{\bm u}^*(\bm k) \sdot \widehat{(\bm u \times \bm \omega)}_{\bm k}$ ($^*$ denotes the complex conjugate) as $\Pi(k) \equiv - \sum_{k' \leq k} T(k')$, from which we compute the maximum downscale energy flux as $\Pi_{\max} \equiv \max[\Pi(k)]$.
as 
\begin{align}
\Pi(k) &\equiv - \sum_{k' \leq k} T(k') \quad \text{with} \\
T(k) &\equiv \sum_{k \leq |\bm k| < k+1} \widehat{\bm u}^*(\bm k) \sdot \widehat{(\bm u \times \bm \omega)}_{\bm k}
\end{align}
the non-linear energy transfer term ($^*$ denotes the complex conjugate), from which we compute the maximum downscale energy flux as $\Pi_{\max} \equiv \max[\Pi(k)]$. We characterise the energy cascade flux by its maximum value $\Pi_{\max}$ since the functional form of $\Pi(k)$ in the inertial-range follows $\Pi(k) \propto \Pi_{\max}(1-\alpha (k\eta)^{4/3})$ for statistically steady turbulence assuming an %Kolmogorov 
energy spectrum $E(k) \propto k^{-5/3}$, where $\alpha$ is a numerical constant and $\eta\equiv (\nu^3/\varepsilon)^{1/4}$ is the Kolmogorov microscale \cite{Ishihara2009,VOS2014}.

Two sets of dimensionless control parameters for the simulations are defined based on the forcing amplitude and large scale turbulence statistics and characterise the turbulence field used as the initial condition. 
The forcing Reynolds and Rossby numbers are given by $Re_F= U/(k_{min}\nu)$ and by $Ro_F = Uk_{min}/(2\Omega)$, respectively, where $U = (f_0/k_{min})^{1/2}$ and $f_0$ is the forcing amplitude. %and $k_{min}$ the smallest wavenumber in the simulation.}
From these definitions $Re_F^2$ is essentially the forcing Grashof number and $Ro_F$ the ratio of the rotation period $\tau_w \propto \Omega^{-1}$ to the %forcing time scale $\tau_{F} = (Uk_{min})^{-1}$.
 turnover time at the forcing scale $\tau_{f} = (Uk_{min})^{-1}$.
% Can we really show that this is a physical time scale?
%
The turbulence Reynolds and Rossby numbers are given by $Re_L=u' L/\nu$ and $Ro_L = u'/2\Omega L$, respectively, where $u' \equiv \sqrt{2/3K}$ is the root-mean-square of the fluctuating velocity. 
For convenience we also define the Taylor microscale based Reynolds number $Re_{\lambda}=u' \lambda/\nu$ where $\lambda\equiv \sqrt{15\nu u'^2/\varepsilon}$ is the Taylor microscale.
Note that $Re_F$ and $Ro_F$ are control parameters that they do not require knowledge of the solution to be evaluated whereas $Re_L$, $Ro_L$ and $Re_\lambda$ are observables and cannot be determined \emph{a priori}. The summary of the control parameters of our DNS and the resulting turbulence Reynolds and Rossby numbers for the initial condition are listed in Table \ref{tbl:dnsparam}.
%

%%%%%%%%%%%%%%%%%%%%%%%%%%%%%%%%%%%%%%%%%%%%%%%%%%%%%%%%%%%%%%%%%%%%%%%%%%%%%%%%%%%%%%%%%%%%%%%%%%%
\begin{table}[!ht]
  \caption{Numerical parameters of the DNS. Note that the values for the resulting turbulence Reynolds and the Rossby numbers are given for the steady state used as initial condition for the decay. Dataset \# 14 was obtained with the numerical code used in Ref. \cite{VOS2014}.}
  \label{tbl:dnsparam}
%   \begin{ruledtabular}
%  \centering
\resizebox{0.48\textwidth}{!}
{
    \begin{tabular}{*{9}{c}}
     \hline
     \hline
     \# & $Re_F$ & $\quad Ro_F \quad$ & $Re_L$ & $\quad Re_\lambda \quad$ & $Ro_L$ & $\quad \Omega \quad$ & $\quad \nu \quad$ & $\quad N \quad$ \\
    \hline
      1 & 200  & 5.0  & 60   & 36  & 2.5 & 0.1  & $5.0 \times 10^{-3}$ & 128  \\
      2 & 200  & 0.5  & 250  & 115 & 0.2 & 1.0  & $5.0 \times 10^{-3}$ & 128  \\
      3 & 667  & 10.0 & 180  & 73  & 6.3 & 0.05 & $1.5 \times 10^{-3}$ & 256  \\
      4 & 667  & 5.0  & 175  & 72  & 3.0 & 0.1  & $1.5 \times 10^{-3}$ & 256  \\
      5 & 667  & 1.0  & 220  & 80  & 0.6 & 0.5  & $1.5 \times 10^{-3}$ & 256  \\
      6 & 667  & 0.5  & 940  & 287 & 0.2 & 1.0  & $1.5 \times 10^{-3}$ & 256  \\
      7 & 2000 &  $\infty$   & 500  & 130 & $\infty$  & 0.0  & $5.0 \times 10^{-3}$ & 512  \\
      8 & 2000 & 5.0  & 500  & 130 & 3.3 & 0.1  & $5.0 \times 10^{-3}$ & 512  \\
      9 & 2000 & var  & 500  & 130 & 3.3 & var  & $5.0 \times 10^{-3}$ & 512  \\
      10 & 2000 & 1.0  & 615  & 144 & 0.6 & 0.5  & $5.0 \times 10^{-3}$ & 512  \\
      11 & 2000 & var  & 615  & 144 & 0.6 & var  & $5.0 \times 10^{-3}$ & 512  \\
      12 & 2000 & 0.5  & 2410 & 414 & 0.2 & 1.0  & $5.0 \times 10^{-3}$ & 512  \\
      13 & 4545 & 5.0  & 1150 & 200 & 3.0 & 0.1  & $2.2 \times 10^{-4}$ & 1024 \\
      14 & -        & $\infty$      & 924  & 173 &$\infty$  & 0.0  & $1.5 \times 10^{-3}$ & 1024 \\
     \hline
     \hline
   \end{tabular}
}
%  \end{ruledtabular}
\end{table}
%%%%%%%%%%%%%%%%%%%%%%%%%%%%%%%%%%%%%%%%%%%%%%%%%%%%%%%%%%%%%%%%%%%%%%%%%%%%%%%%%%%%%%%%%%%%%%%%%%%

According to Ref. \cite{davidsonetal12ten}, it is common practice in turbulence laboratory experiments to force briefly at Rossby numbers $Ro_L > 1$, and then let $Ro_L$ drift down to $Ro_L \sim 1$ as the energy of the turbulence decays. Trying to perform numerical simulations that would be useful to laboratory experiments, we followed this approach to perform our DNS. As we shall see this approach is rather different to the simulations which often have imposed $Ro \ll 1$ as an initial condition.

\begin{figure}[ht]
\includegraphics[width=85mm]{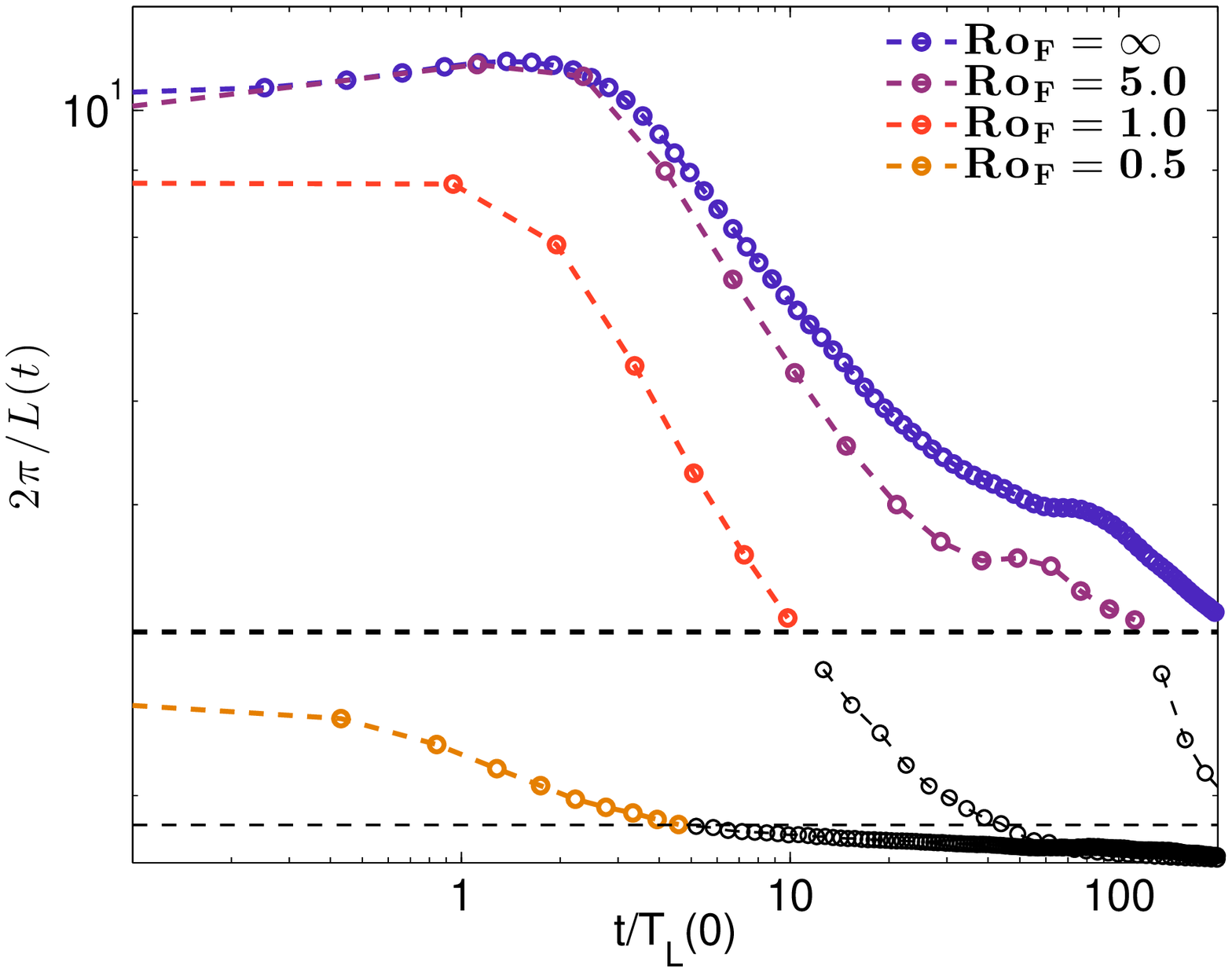}
\caption{
Development of the ratio of the box-size to the integral length-scale, $2\pi/L$, throughout the decay for various  control Rossby numbers $Ro_F$ at $Re_F=2000$. The abscissas are normalised by the initial eddy turnover time. The data corresponding to $2\pi/L<4$ (or $2\pi/L<2.9$ for the strong rotation case, $Ro_F=0.5$) are denoted using black thin markers and dashed lines. %\red{Pedro: Are you not plotting the time series of the integral scales that I have outputed? -- No, as I wrote in the e-mail on May 20th, I got all the data from the spectra for simplicity, clarity and to be able to compare with my previous simulations.}
%Growth of the integral length-scale $L$ over time (normalised by the initial eddy turnover time) for (a) various initial turbulent Rossby numbers $Ro_L(0)$ at $Re_F=2000$ and (b) various initial turbulent Reynolds numbers $Re_L(0)$ at $Ro_F=5.0$.
}
\label{fig:Lintplots}
\end{figure}

To avoid biasing our analyses with data that may have non-negligible confinement effects due to the periodic box size we only consider data points where the integral length-scale is smaller than $1/4$ of the box size ($2\pi/L \gtrsim 4$; \cite{BKR99}), except for the strongly rotating cases (datasets \# 2, \# 6 and \# 12) where we alleviate this constraint to $2\pi/L \gtrsim 2.9$ (see Fig. \ref{fig:Lintplots}). Nevertheless, we include the remaining data in the figures, but distinguish them using black markers and thin dashed lines.

\section{\label{sec:results}Results}

The presented  data complements the numerical and experimental data available in the literature in two fundamental aspects. 
Firstly, we use forced statically steady state turbulence with the desired Rossby number as an initial condition rather than a randomised velocity field. 
This approach allows us to reproduce the conditions for non-equilibrium turbulence dissipation \cite{VOS2014} and assess for the first time whether it also occurs in mildly rotating turbulence.
It also guarantees that the turbulence is fully developed - in the sense of a fully developed %kinetic 
energy cascade - from the very start of the decay in contrast to the standard approach where the first couple of eddy turnover times of the decay are biased by the development of the non-linear interactions.   
This allows us to consider the data from the very start of the decay, where the non-equilibrium dissipation behaviour is manifested, but comes at the price of requiring a converged forced run for every decay simulation.
Secondly, in addition to simulations with a constant rotation rate and thus decreasing Rossby number during decay (i.e. an increasing influence of the background rotation) we also perform decaying simulations with a constant Rossby number by varying the rotation rate. 
This, in turn, allows us to study the decay of turbulence subjected to rotation within the same rotating turbulence regime, i.e. maintaining the same ratio of the rotation period $\tau_w \propto \Omega^{-1}$ to the eddy turnover time $T_L = L/u'$ throughout the decay, and report differences to the standard approach of fixing $\Omega$ and thus  straddling multiple rotating turbulence regimes during the decay (since %$L/u'$ 
$T_L$ can increase by multiple orders of magnitude).

\begin{figure}[ht]
\includegraphics[width=85mm]{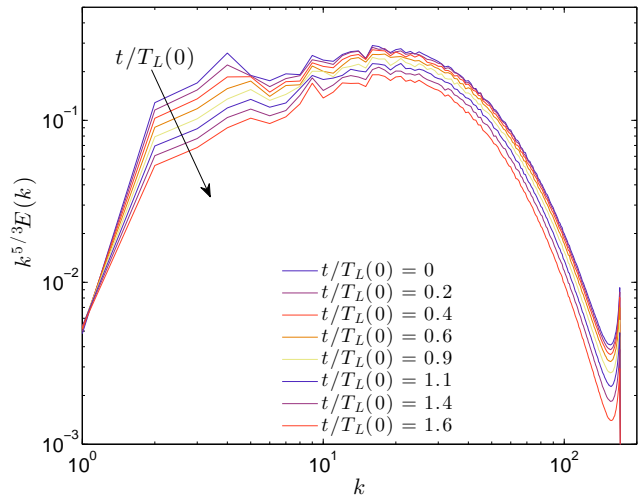}
\caption{Energy spectrum compensated by $k^{5/3}$ for the initial instants of the turbulence decay for  $Ro_F=\infty$ and $Re_F=2000$.} 
\label{fig:Spectrum}
\end{figure}%

\begin{figure}[ht]
\includegraphics[width=85mm]{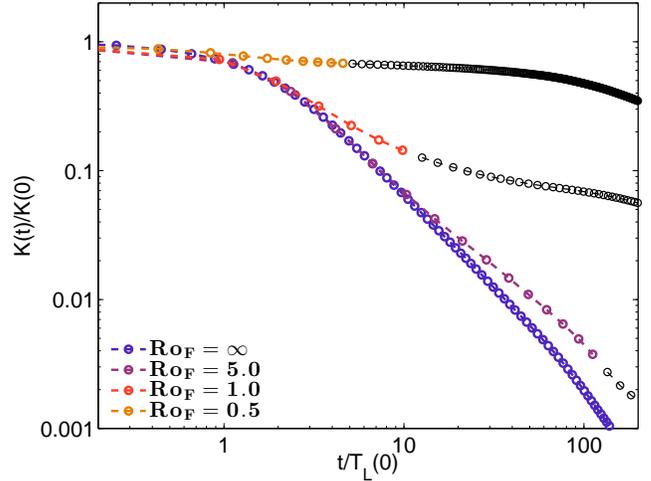}
\caption{Decay of %kinetic 
energy over time (normalised by the initial eddy turnover time) for various  control Rossby numbers $Ro_F$ at $Re_F=2000$.}
\label{fig:timeplots}
\end{figure}%

\begin{figure}[ht]
\includegraphics[width=85mm]{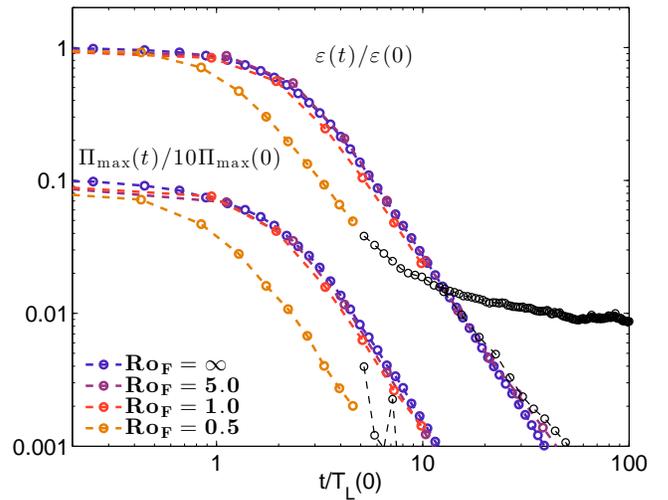}
\caption{Decay of the turbulent %kinetic 
energy dissipation $\varepsilon$ and the energy cascade flux $\Pi$  for different control Rossby numbers $Ro_F$ for $Re_F=2000$.}
\label{fig:Rossbyplots}
\end{figure}

\begin{figure}[ht]
\includegraphics[width=85mm]{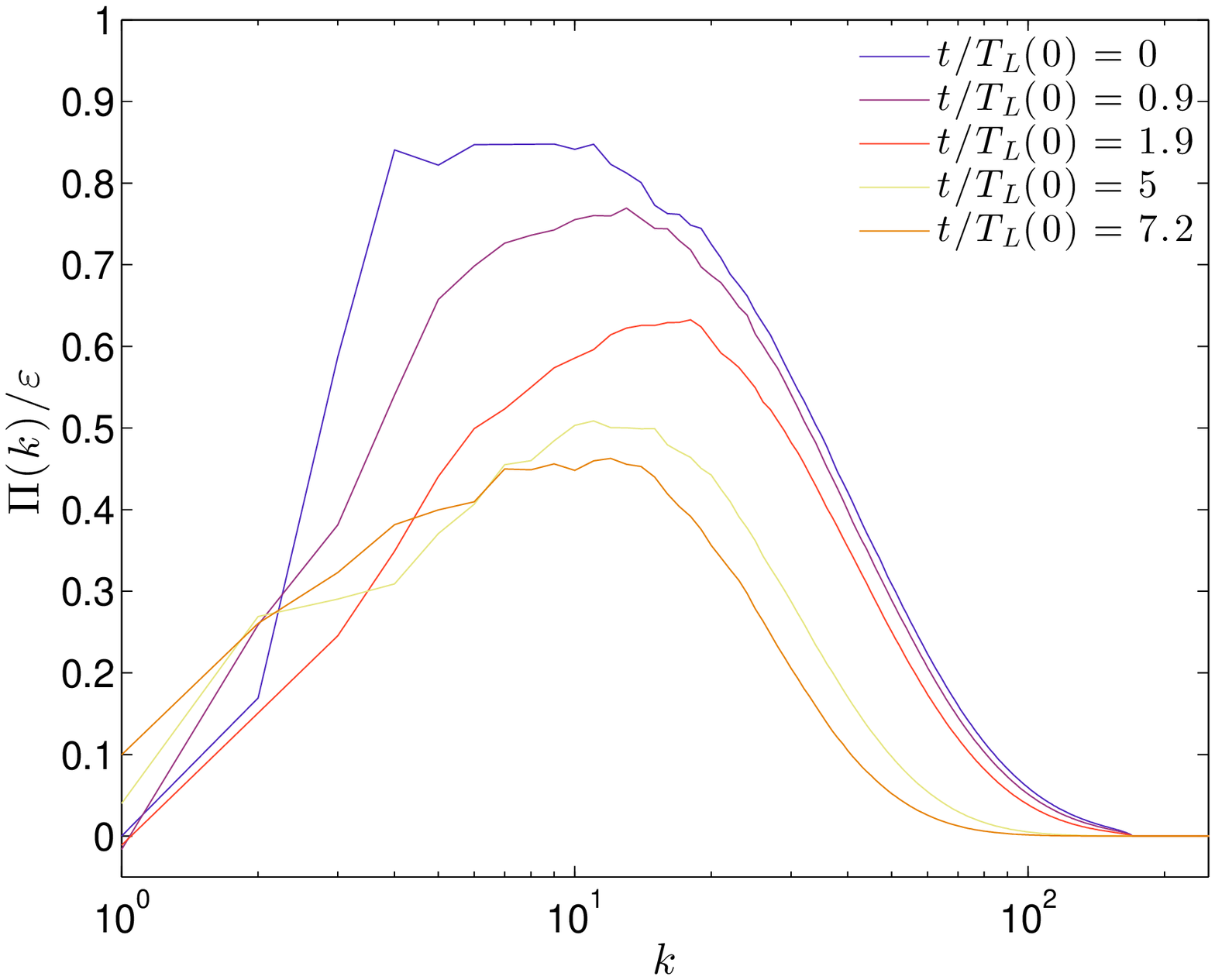}
\caption{Energy cascade flux spectrum $\Pi(k)$ for various snapshots throughout the decay for dataset \# 10 ($Re_F=2000$, $Ro_F=1.0$).}
\label{fig:fluxspectrum}
\end{figure}

\subsection{Temporal evolution}
We start by presenting the temporal evolution of the turbulence statistics that will be used to show that in rotating turbulence there is also evidence of non-equilibrium dissipation scalings and of the imbalance $\Pi(k) \neq \varepsilon$ throughout the decay.
The statistics of interest are the time-series of the integral scale $L$ (Fig. \ref{fig:Lintplots}), the kinetic 
energy $K$ (Fig. \ref{fig:timeplots}), the dissipation rate $\varepsilon$ and the maximum energy %cascade 
flux $\Pi_{max}$ (Fig. \ref{fig:Rossbyplots}) %as a function of the control Rossby number 
for a range of the control parameter $Ro_F$.
As noted in \S \ref{sec:dns}, box-turbulence simulations can be hindered by confinement effects if the integral scale is not sufficiently smaller than the box size. 
The situation worsens for decaying box-turbulence simulations since the integral scale generally grows throughout the decay and thus the effects of confinement are progressively larger. 
We chose $2\pi/L\approx 4$ \cite{BKR99} as our cutoff beyond which the confinement effects may no longer be negligible (Fig. \ref{fig:Lintplots}). 
For rotating turbulence, the integral scale tends to increase and grow faster during decay for smaller Rossby numbers (i.e. larger $\Omega$),  arguably due to the effect of the background rotation on the %kinetic 
energy cascade \cite{Moisy2006,mininni11,Cambon2014}, and therefore the confinement tends to deteriorate (Fig. \ref{fig:Lintplots}). 
Note that at the start of the decay the integral scale decreases ($2\pi/L$ increases) for the first one or two turnover  times before growing throughout the remaining decay. 
We observed this behaviour for most of our simulations, except the strongly rotating cases. 
To the best of our knowledge, this behaviour has not been reported before and renders some discussion. 
Given that $L \equiv 3\pi/(4 K) \sum_{k} E(k)/k$, the low wavenumber part of the spectrum has a large influence on the numerical value of $L$ and thus a decrease in $L$ during a decrease in $K$ implies that the smaller wavenumbers are loosing %kinetic 
energy faster than the larger wavenumbers (cf. Fig. \ref{fig:Spectrum}). 
%
% Based on the data presented in Fig. \ref{fig:Spectrum}, 
A plausible explanation is the adjustment of the low wavenumber part of the spectrum to the cessation of external forcing - noticeable up to $t/T_L(0)\lesssim 0.4$ in Fig. \ref{fig:Spectrum}. 

In turn the %kinetic 
energy %monotonously 
decreases monotonically throughout the decay at a rate which depends on the initial Rossby number (Fig. \ref{fig:timeplots}), which is consistent with the numerical and experimental data in the literature \cite{Moisy2006,Davidson2008,mininni11}.
Given that in freely decaying homogeneous turbulence $d K/dt = -\varepsilon$, this is a direct consequence of the faster decrease in the energy dissipation rate $\varepsilon$, which is a consequence of (or the cause for) the dampening of the non-linear energy %cascade 
flux $\Pi(k)$ (Fig. \ref{fig:Rossbyplots}).
The energy flux spectra for various instances throughout the decay are shown in Fig. \ref{fig:fluxspectrum} for a dataset with moderately strong background rotation ($Ro_F=1.0$ and $Ro_L$ decreases from 0.5 at the start of the decay to 0.1 when $2\pi/L < 4$;
%\textcolor{magenta}{ 
this dataset is included in Figs \ref{fig:Lintplots}, \ref{fig:timeplots} and \ref{fig:Rossbyplots}).
%}
%
%\textcolor{magenta}{
It can be noted that the maximum value of the normalised flux spectrum $\Pi(k)/\varepsilon$ decreases as time progresses and that no upscale energy flux at low wavenumbers occurs even at later times where the Rossby number is moderately low, $Ro_{L}\approx 0.1$. 
Fig. \ref{fig:fluxspectrum} also illustrates what we mean by mildly rotating turbulence - turbulence where the background rotation has a significant effect on the energy cascade but not strong enough to induce an upscale energy flux and/or a quasi-2D flow.
%}

\subsection{Decaying non-rotating turbulence}

Having turbulence modelling in mind, in the following discussion we chose to consider the dimensionless dissipation and energy flux parameters and how they may scale with large scale variables,
\begin{equation}
 C_{\varepsilon} \equiv \varepsilon L/ u'^{3} \quad \text{and} \quad C_{\Pi}\equiv \Pi_{\max} L/ u'^{3},
\end{equation}
respectively, without attempting to infer whether statistics decomposed along axes parallel and perpendicular to the axis of the rotation could improve the scalings (see e.g., Ref. \cite{mininnietal09}).
 
Prior to discussing our results concerning the scaling behaviour of these quantities in rotating turbulence, it is useful to review the recent developments for non-rotating turbulence to have it as benchmark. 
In laboratory experiments of grid-generated decaying non-rotating turbulence it is widely accepted that far from the grid %the dissipation is proportional to the kinetic energy over the eddy furnover time,
$\varepsilon \propto u'^{3} / L$ %K/T_L$ %i.e.
or $C_{\varepsilon} \approx \mathrm{constant}$ as long as the Reynolds number of the decaying turbulence remains moderately large, typically at least above $Re_{\lambda}\approx 100$. 
Although until recently the scaling of the cascade flux had not been measured and the Reynolds number of the DNS data were insufficiently large it was believed that in that same region $C_{\Pi}\approx C_{\varepsilon} \approx \mathrm{constant}$. 
Recently, however, three interesting findings have been reported for both laboratory experiments and numerical simulations. 
Firstly, it was found that upstream or %before the region where  $C_{\varepsilon} \approx \mathrm{constant}$, 
after the steady-state region (i.e. $C_\varepsilon \approx C_\Pi$), there is a region where $C_{\varepsilon} \propto Re_0/Re_L \neq  \mathrm{constant}$, denoted as a non-equilibrium dissipation region \cite{VV2012,VOS2014,VV2015} (where $Re_0$ is a global Reynolds number of the flow such as our control Reynolds number based on the forcing $Re_F$ or a mesh Reynolds number for grid turbulence experiments). 
Secondly, it was found that in the further downstream region or later in time where $C_{\varepsilon} \approx \mathrm{constant}$, the dissipation was roughly twice the non-linear flux $\Pi_{\max}$ (i.e. $C_{\varepsilon} \approx 2C_\Pi$). %and that the non-equilibrium region covered the `transition' between %a steady-state, where $\varepsilon \approx \Pi_{\max}$, and the further downstream or later in time region where $\varepsilon \approx 2\Pi_{\max}$ 
%\vd{the steady-state (i.e $\varepsilon \approx \Pi_{\max}$) and the region where $\varepsilon \approx 2\Pi_{\max}$}
%\cite{VV2015,VOS2014} (note that for the wind tunnel experiments the statistically steady point occurs at the start of the decay when $Udu'^2/dx = 0$, where $U$ is the mean flow velocity, see  Ref. \cite{VV2015}). 
%
Finally, it was found that $C_{\Pi}$ exhibits much smaller variations and can be considered to a first approximation to being constant throughout the decay, contrary to what is observed for $C_{\varepsilon}$. 

\begin{figure}[!ht] 
\includegraphics[width=85mm]{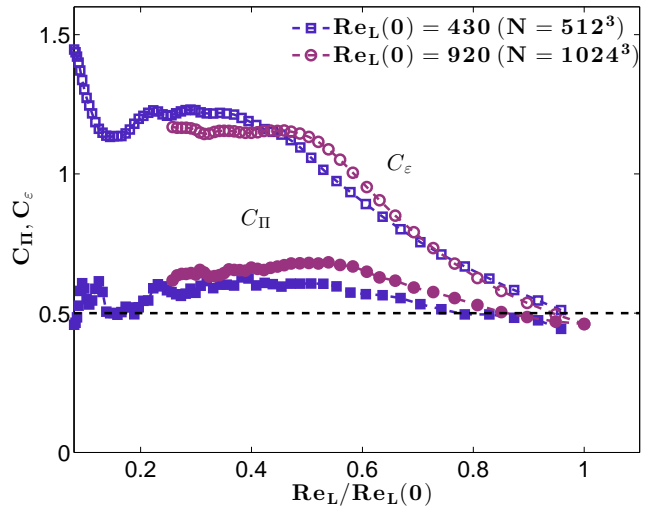}
\caption{Normalised turbulent energy dissipation $C_{\varepsilon}$ and non-linear energy cascade flux $C_{\Pi}$ versus the turbulent Reynolds number $Re_{L}$ for non-rotating decaying simulations starting from a statistically steady forced state.}
\label{fig:CepsCpivsRoHydro}
\end{figure}

In Fig. \ref{fig:CepsCpivsRoHydro} we present data for $C_{\Pi}$ and $C_{\varepsilon}$ from two DNSs of decaying %periodic-box 
non-rotating turbulence which are consistent with the above mentioned findings.
The steady state corresponds to the initial point where $Re_{L}/Re_L(0)=1$ and $C_{\Pi}\approx C_{\varepsilon}$. 
As the turbulence decays, the Reynolds number decreases and the data show that $C_{\varepsilon}$ increases from its steady state value around 0.5 until reaching a plateau starting around $Re_L/Re_L(0)\approx0.5$ where it takes a numerical value of order 1.
This is the denoted non-equilibrium dissipation region exhibiting a clear departure from $C_{\varepsilon}\approx \mathrm{constant}$, contrasting with the behaviour $C_{\Pi}$ which exhibits a lesser variation. 
As the turbulence continues to decay the Reynolds number will eventually decrease to a point where low Reynolds number effects will be non-negligible and $C_{\varepsilon}$ will depart from the plateau with the numerical value of order 1, whereas $C_{\Pi}$ remains approximately constant  (see e.g., Ref. \cite{McComb2010} for a review on the low Reynolds number behaviour of $C_{\varepsilon}$ and data supporting $C_{\Pi}\approx \mathrm{constant}$ for low Reynolds numbers).
%
%For our data this occurs around $Re_L/Re_L(0)\approx0.2$. 
%

Note that the data presented in Fig. \ref{fig:CepsCpivsRoHydro} was obtained with two distinct numerical codes. The lower Reynolds number data simulated with $N=512^3$ collocation points was obtained with the numerical code used for the remaining rotating turbulence simulations, whereas the $N=1024^3$ data was obtained with the numerical code used in Ref. \cite{VOS2014}. 
Both numerical codes employ a pseudo-spectral method, but the forcing strategies for the steady state simulations serving as initial conditions for the decay are quite different. %(see \S \ref{sec:dns}). 
%
%\vd{The difference is in the two non-helical random forcing strategies that were used to generate the initial conditions. 
For more details please refer to Refs. \cite{VOS2014,dt16}. %is that the latter is an impulsive force uncorrelated with the velocity field and thus allows to set \emph{a priori} the level of dissipation in the steady state, whereas the former specifies the amplitude of the forcing.

It is thus reassuring to note that although there are quantitative differences, the qualitative behaviour of $C_{\Pi}$ and  $C_{\varepsilon}$ is quite similar.

\subsection{Decaying rotating turbulence}

\begin{figure}[!ht]
 \begin{subfigure}{0.49\textwidth}
   \includegraphics[width=\textwidth]{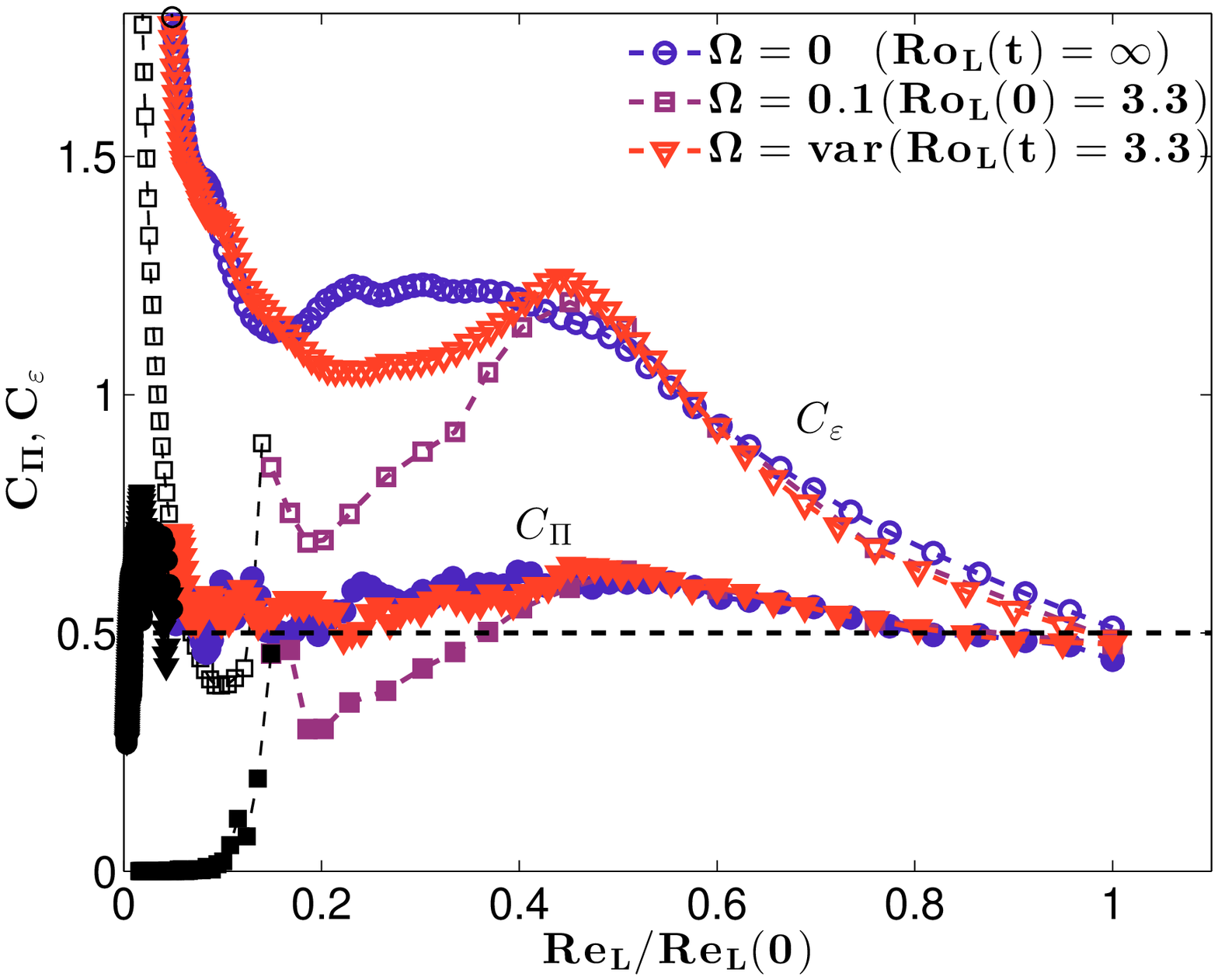}
   \caption{}
   \label{fig:}
 \end{subfigure}
 \begin{subfigure}{0.49\textwidth}
   \includegraphics[width=\textwidth]{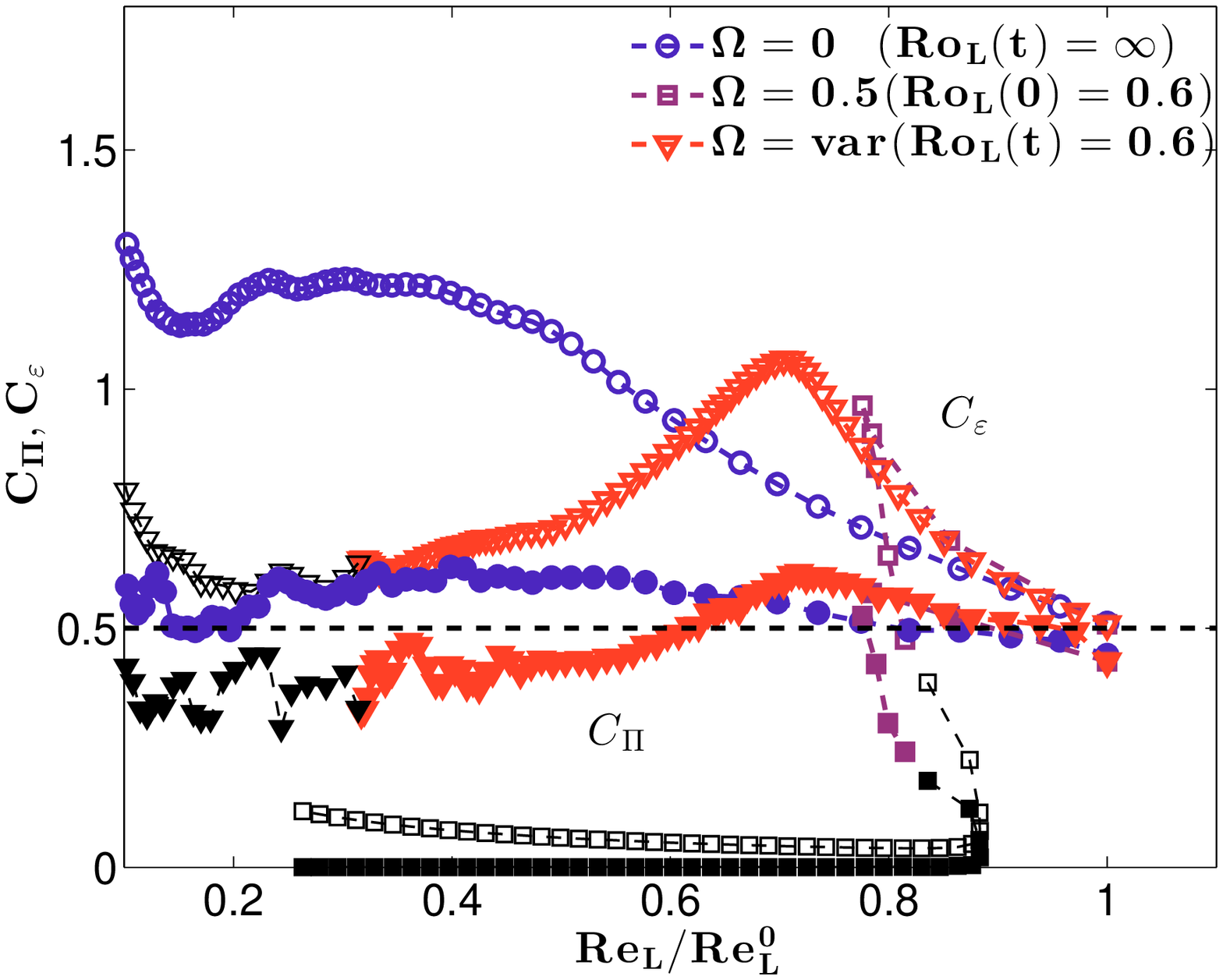}
   \caption{}
   \label{fig:}
 \end{subfigure}
\caption{Normalised turbulent %kinetic 
energy dissipation $C_{\varepsilon}$ and non-linear energy cascade flux $C_{\Pi}$ versus the turbulent Reynolds number $Re_{L}$ for rotating and non-rotating decaying simulations starting from a statistically steady forced state. For the rotating cases we compare both constant rotation rate $\Omega$ (and varying $Ro_L$) and constant turbulent Rossby number $Ro_L$ (achieved by varying $\Omega$) simulations. }
\label{fig:CepsCpivsRe}
\end{figure}

\begin{figure}
\includegraphics[width=85mm]{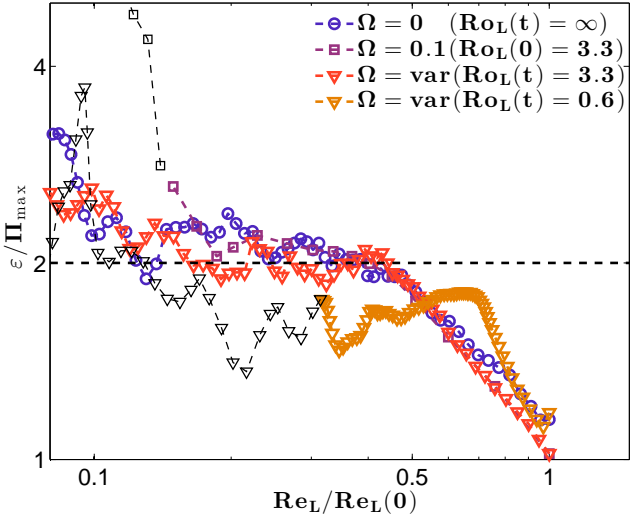}
\caption{Ratio between the turbulent %kinetic 
energy dissipation $\varepsilon$ and the maximum non-linear energy cascade flux $\Pi_{\max}$ versus the turbulent Reynolds number $Re_{L}/Re_L(0)$ for rotating and non-rotating decaying simulations starting from a statistically steady forced state. }
\label{fig:CepsoverCpivsRe}
\end{figure}

\begin{figure}[!ht]
 \begin{subfigure}{0.49\textwidth}
   \includegraphics[width=\textwidth]{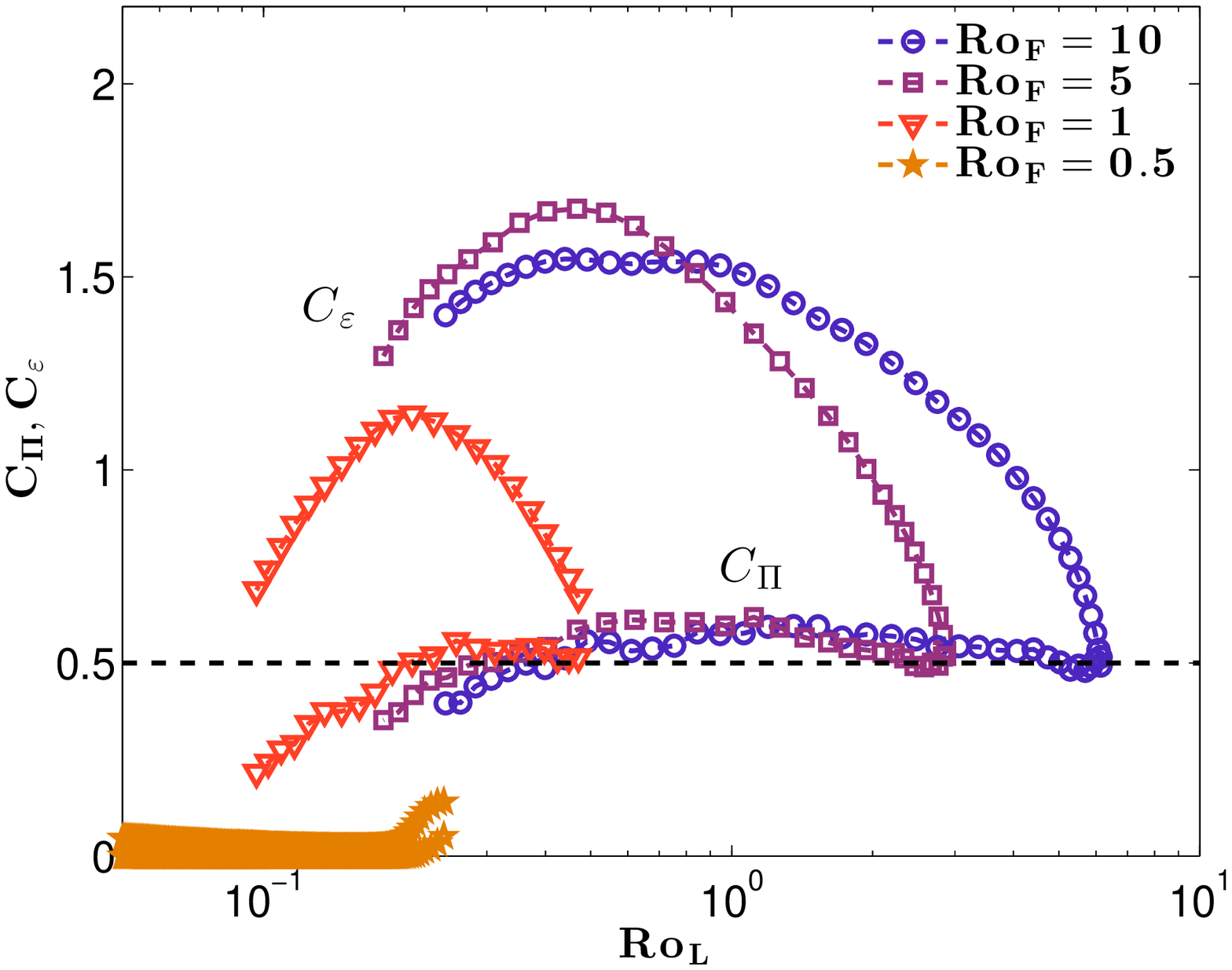}
   \caption{}
   \label{fig:}
 \end{subfigure}
% \vspace{2.0cm}
 \begin{subfigure}{0.47\textwidth}
   \includegraphics[width=\textwidth]{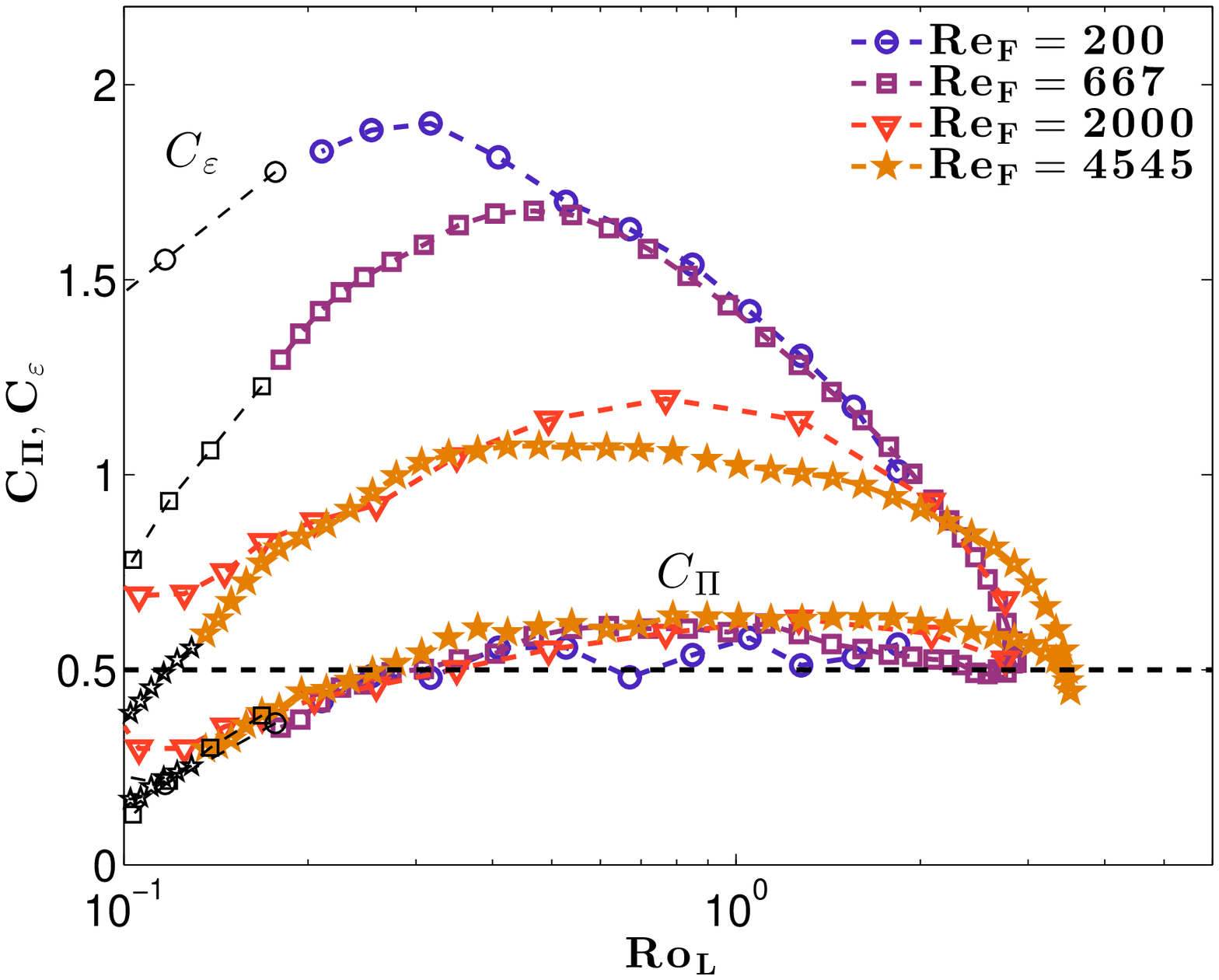}
   \caption{}
   \label{fig:}
 \end{subfigure}
\caption{Normalised turbulent %kinetic 
energy dissipation $C_{\varepsilon}$ and non-linear energy cascade flux $C_{\Pi}$ versus the turbulent Rossby number $Ro_{L}$ for the rotating decaying simulations starting from a statistically steady forced state for (a) various control Rossby numbers $Ro_F$ at a fixed $Re_F=667$ and (b) various control Reynolds numbers $Re_F$ at a fixed $Ro_F=5.0$.}
\label{fig:CepsCpivsRo}
\end{figure}

Turning now to the decaying rotating turbulence, %subjected to rotation, 
it is clear that the same qualitative departure between $C_{\Pi}$ and  $C_{\varepsilon}$ occurs from the start of the decay for both cases of weak ($Ro_L \approx 3.3$) and %strong 
stronger rotation ($Ro_L \approx 0.6$) and for both fixed and varying rotation rates (cf. Figs. \ref{fig:CepsCpivsRe} and \ref{fig:CepsCpivsRo}).
However, rather than reaching a plateau, $C_{\varepsilon}$ reaches a maximum value and decreases afterwards (see Figs. \ref{fig:CepsCpivsRe}a and \ref{fig:CepsCpivsRe}b).
Interestingly, this appears to be directly associated with the behaviour of $C_{\Pi}$ and how %it is 
the non-linear interactions are affected by the background rotation. 
For the runs with fixed background rotation rate, and therefore decreasing $Ro_L$ as the turbulence decays, the associated dampening of the energy cascade leads to a diminishing value of $C_{\Pi}$ which occurs progressively for the run with weaker background rotation (Fig. \ref{fig:CepsCpivsRe}a) and very %aggressively 
abruptly for the run with higher background rotation (Fig. \ref{fig:CepsCpivsRe}b). 
For the runs with fixed $Ro_L$ throughout the decay (i.e. varying $\Omega$), it is clear that the effect of the background rotation on the cascade  leads to a reduced variation in the numerical value of $C_{\Pi}$ throughout the decay.  For the weaker rotation ($Ro_L\approx 3.3$) the behaviour of $C_{\Pi}$ is almost identical to the non-rotating case (Fig. \ref{fig:CepsCpivsRe}a), whereas for the stronger rotation ($Ro_L\approx 0.6$), the normalised energy flux reaches a plateau around $C_{\Pi}\approx 0.4$ (Fig. \ref{fig:CepsCpivsRe}b).
Interestingly, it appears that the behaviour of $C_{\varepsilon}$ beyond the initial increase appears to be dictated by the behaviour of $C_{\Pi}$ and %, to the trained eye, 
the two seem to be proportional. 
Indeed, by plotting the ratio between $\varepsilon$ and $\Pi_{\max}$ it can be seen that there is a period where $\varepsilon/\Pi_{\max} \approx \mathrm{constant}$ (with a constant around 2 or slightly lower for the stronger rotation case, which may nevertheless be due to confinement effects)
after a transition region from the initial steady state where $\varepsilon \approx \Pi_{\max}$, similar to what is observed for  non-rotating turbulence (cf. Fig. \ref{fig:CepsoverCpivsRe}).
As noted for non-rotating turbulence, as the turbulence continues to decay, the small Reynolds number effects become non-negligible and there is a monotonous increase in the ratio %between $\varepsilon$ and $\Pi_{\max}$. 
$\varepsilon/\Pi_{\max}$. 

%Finally, 
In Fig. \ref{fig:CepsCpivsRo}, we show the behaviour of $C_{\Pi}$ and  $C_{\varepsilon}$ against the turbulent Rossby number $Ro_L$ for various control $Ro_F$ when $Re_F = 667$ (Fig. \ref{fig:CepsCpivsRo}a) and for various control $Re_F$ when $Ro_F = 5.0$ (Fig. \ref{fig:CepsCpivsRo}b).
% %
% \begin{figure}
% \begin{minipage}[c]{0.49\textwidth}
%  \centering
%  \begin{lpic}[]{Plots/Paper_CepsoverCpivsRoL_b(85mm)}
%  \lbl{2,123;(a)} 
%  \end{lpic}
% \end{minipage}
% \begin{minipage}[c]{0.49\textwidth}
%  \centering
%   \begin{lpic}[]{Plots/Paper_CepsoverCpivsRoL(85mm)}
%  \lbl{2,123;(b)} 
%  \end{lpic}
%  \end{minipage}
% \caption{Ratio between the turbulent energy dissipation $\varepsilon$ and the maximum non-linear energy cascade flux $\Pi_{\max}$ versus the turbulent Rossby number $Ro_{L}$ for rotating decaying simulations starting from a statistically steady forced state for (a) various control Rossby numbers $Ro_F$ at a fixed $Re_F=667$ and (b) various control Reynolds numbers $Re_F$ at a fixed $Ro_F=5.0$.}
% \label{fig:CepsoverCpivsRo}
% \end{figure}
%
Interestingly, it appears that $C_{\Pi}$ is roughly constant with a numerical value around 0.5 for %Rossby numbers larger than $Ro_L \approx 0.5$ 
$Ro_L \gtrsim 0.3$ and decreases for smaller $Ro_L$. %without a significant impact 
Note that $C_{\Pi}$ is almost independent of the initial Rossby and  Reynolds numbers as long as the initial Rossby number is sufficiently large to allow a fully turbulent flow for a given Reynolds number \cite{Alexakis2015}.
In turn, the behaviour of $C_{\varepsilon}$ for the various initial Rossby numbers is also qualitatively similar to that discussed above, i.e. presenting the initial ascending departure from $C_{\varepsilon} \approx C_{\Pi}$ followed by a descent which 
%is roughly proportional to the decrease in $C_{\Pi}$. 
can be attributed to the effect of the background rotation on the energy cascade that is depicted as a descrease in $C_{\Pi}$.
% Note that the decrease in $C_{\varepsilon}$ is more evident solely because it has a larger numerical value, but by plotting the ratio between the two as in Fig. \ref{fig:CepsoverCpivsRe} it becomes clear that they are proportional.

Lastly, the behaviour of $C_{\varepsilon}$ for increasingly larger %control Reynolds numbers 
$Re_F$ is such that the maximum value reached decreases 
%as well as the extent of the region where $C_{\varepsilon}$ is inversely proportional to $Ro_L$ 
(cf. Fig. \ref{fig:CepsCpivsRo}b).  
For even larger Reynolds numbers, it may be the case that the behaviour of $C_{\varepsilon}$ becomes Reynolds number independent with a functional form close to that indicated by our largest Reynolds number dataset in Fig. \ref{fig:CepsCpivsRo}b, but one cannot preclude the hypothesis that the departure of $C_{\varepsilon}$ from $C_{\Pi}$ will further decrease and eventually $C_{\varepsilon} \approx C_{\Pi}$, indicating that this behaviour is a finite Reynolds number effect which vanishes at very large $Re$.
%%
%\red{Our data are not able to provide such an answer, but the reader is referred to the discussion in Ref. \cite{VOS2014} where a compilation of data on the imbalance between $\varepsilon$ and $\Pi_{\max}$ for non-rotating hydrodynamic turbulence suggests its persistence up to $Re_{\lambda}\sim\mathcal{O}(10^5)$. 
%%Pedro: how relevant is this here?
%%
%Dedicated laboratory and numerical experiments at larger Reynolds numbers are thus necessary to settle this debate.}
%%Pedro: again alexakis15

%
\section{\label{sec:conclusions}Conclusions}
% \red{ADD: A comment in the conclusions about $C_\Pi = const$}

% \vd{In this study, we investigate the existence of significant imbalances between the energy flux $\Pi(k)$ and the dissipation rate $\varepsilon$ by means of DNS of decaying, mildly rotating turbulent flows, i.e. flows where the background rotation has a significant effect on the energy cascade but not strong enough to induce an upscale energy flux. Our decaying flows are subjected to different rotation rates and we use a statistically steady and fully developed rotating turbulence field as an initial condition.}

Decaying turbulence subjected to mild background rotation exhibits similar imbalances between the energy %cascade 
flux $\Pi_{\max}$ and the %kinetic 
energy dissipation rate $\varepsilon$ as recently reported for laboratory and numerical experiments of freely decaying non-rotating turbulence \cite{VOS2014,VV2015}. 
In close resemblance to non-rotating turbulence, the ratio %between $\varepsilon$ and $\Pi$
$\varepsilon/\Pi_{\max}$ increases from unity at the start of the decay, if the initial condition is statistically steady turbulence at sufficiently large Reynolds number, up to a value around 2 where it exhibits a plateau which ceases when the turbulence has decayed to a point where low Reynolds number effects become predominant.

At the initial stage of the decay the dimensionless parameter $C_\Pi$ is approximately contast (i.e. $\Pi_{max} \propto u'^3/L$), while $C_\varepsilon$ increases up to a maximum value. In contrast to non-rotating turbulence, we find that %decaying and mildly rotating turbulence 
$C_\varepsilon$ does not exhibit a region where $C_{\varepsilon} \approx \mathrm{constant}$ (i.e. $\varepsilon$ does not scale as $u'^3/L$).
This appears to be related to the fact that $C_{\Pi}$ tends to decrease after $C_{\varepsilon}$ reaches a maximum value as the Rossby number and the Reynolds number decrease. %The latter 
The decrease in $C_\Pi$ is commonly attributed to the dampening of the non-linear energy cascade caused by the background rotation.
We demonstrate this by introducing simulations with fixed turbulent Rossby number. In this case, it is possible to maintain a consistent effect of the rotation throughout the decay, %and decrease the variation in $C_{\Pi}$ which also decreases the variation in $C_{\varepsilon}$.
which reduces the variation in $C_\Pi$ and concequently the variation in $C_\epsilon$. 

Our data indicates that $C_{\varepsilon}$ may not tend towards $C_{\Pi}$ as the Reynolds number increases, but 
we are not able to address how this imbalance will behave at larger Reynolds numbers.
However, for non-rotating flows the reader is referred to the discussion in Ref. \cite{VOS2014} and the data compilation on Refs. \cite{Antonia2006,Cambon2012} where the imbalance between $\varepsilon$ and $\Pi_{\max}$ is suggested to persist up to at least $Re_{\lambda}\sim\mathcal{O}(10^5)$, which implies that for the overwhelming majority of engineering applications one cannot neglect this behaviour.
%Pedro: again how relevant is this for rotating flows? 

The fact that $C_\Pi$ remains constant while $C_{\varepsilon}$ exhibits significant variations during decay for the Reynolds and Rossby numbers that we considered implies that a turbulence model, in the spirit of the $K$-$\varepsilon$ model, with an evolution equation for the energy flux instead of the energy dissipation rate would be more robust for the simulation of non-stationary flows ubiquitous in engineering applications. Nevertheless, in order to have a more complete picture, dedicated experiments are required to assess %the non-equilibrium dissipation scaling and 
the imbalance $\Pi_{\max} \neq \varepsilon$ at much larger Reynolds numbers for both rotating and non-rotating turbulent flows that are statistically non-stationary. Until recently these investigations were limited to laboratory experiments but the computational capabilities to perform high Reynolds number numerical experiments is now becoming available. 

\begin{acknowledgements}
VD acknowledges support from the Royal Society and the British Academy of Sciences (Newton International Fellowship, NF140631). The computations were performed on ARC1 and ARC2, part of the High Performance Computing facilities at the University of Leeds, UK. PV acknowledges support from COMPETE, FEDER and Funda\c{c}\~{a}o para a Ci\^{e}ncia e a Tecnologia (grant PTDC/EME-MFE/113589/2009) on an early stage of the work and would like to thank Prof. Carlos B. da Silva for making possible the non-rotating turbulence simulation with $N=1024^3$ collocation points.
\end{acknowledgements}
\bibliography{mybib}
\end{document}